\documentclass[prl,twocolumn,showpacs,showkeys,preprintnumbers,amsmath,amssymb]{revtex4}

\usepackage{graphicx} 
\usepackage{dcolumn}  
\usepackage{bm}       

\begin{document}

\preprint{GST compounds}

\title{Atomistic origins of the phase transition mechanism in Ge$_2$Sb$_2$Te$_5$}  
\author{Juarez L. F. Da Silva,$^1$ Aron Walsh,$^1$ Su-Huai Wei,$^1$ and Hosun Lee$^2$}
\affiliation{$^1$National Renewable Energy Laboratory, 1617 Cole Blvd., 
Golden, CO 80401, USA \\ $^2$Dept. of Applied Physics, Kyung Hee University, Suwon 
446-701, South Korea}
\date{\today}

\begin{abstract}
Combined static and molecular dynamics first-principles calculations are used 
to identify a direct structural link between the metastable crystalline and 
amorphous phases of Ge$_2$Sb$_2$Te$_5$. We find that the phase transition is 
driven by the displacement of Ge atoms along the rocksalt [111] direction from 
the stable-octahedron to high-energy-unstable tetrahedron sites close to the 
intrinsic vacancy regions, which give rise to the formation of local 4-fold 
coordinated motifs. Our analyses suggest that the high figures of merit of 
Ge$_2$Sb$_2$Te$_5$ are achieved from the optimal combination of intrinsic 
vacancies provided by Sb$_2$Te$_3$ and the instability of the tetrahedron 
sites provided by GeTe. 
\end{abstract}
 
\maketitle

Ternary (GeTe)$_m$(Sb$_2$Te$_3$)$_n$ materials, in particular the 
Ge$_2$Sb$_2$Te$_5$ (GST) composition, have been considered as the most natural 
candidates for non-volatile memory applications through exploiting the fast 
and reversible resistance change between a metastable (m-GST) crystalline 
phase (low resistivity) and an amorphous (a-GST) phase (high resistivity).
\cite{Ovshinsky-1968-1450,Yamada-1991-2849,Wuttig-2007-824,Chong-2008-136101} 
However, the mechanism of the phase transition is still under intense debate. 
The existing models,\cite{Kolobov-2004-703,Kolobov-2006-035701,Welnic-2006-56,Hegedus-2008-399} 
have provided a preliminary understanding of the transition mechanism, but 
fail to provide a clear and direct structural link between the m-GST and a-GST 
phases, which play a key role in the understanding of the reservible 
transition at an atomistic level. The m-GST\cite{Yamada-2000-7020,Njoroge-2002-230,Matsunaga-2004-104111,Park-2005-093506,Sun-2006-055507,Wuttig-2007-122,Shportko-2008-653} 
phase crystallizes in a rocksalt-type (RS) structure, in which the Te atoms 
occupy the anion sites and Ge, Sb, and the naturally occurring intrinsic 
vacancies from Sb$_2$Te$_3$ (20\% in GST) occupy the cation sites. It has been 
suggested that a-GST is characterized by the presence of 4-fold coordinated Ge 
atoms,\cite{Kolobov-2004-703,Baker-2006-255501,Akola-2007-235201,Caravati-2007-171906,Hegedus-2008-399,Jovari-2008-035202} 
in which the sum of the occurrences GeTe$_4$, Ge(SbTe$_3$), and Ge(GeTe$_3$) 
is about 66\%.\cite{Jovari-2008-035202} Ge$-$Ge and Ge$-$Sb bonds are found in 
those motifs, which is assumed to be due to disorder effects, since they are 
not present in the crystalline phases. Furthermore, a-GST shows a volume 
expansion of $6-7$\% compared with the m-GST phase,
\cite{Weidenhof-1999-5879,Njoroge-2002-230} and it has a higher energy 
($28-40$~meV/atom) with respect to m-GST.\cite{Kalb-2003-2389} Theoretically, 
first-principles molecular dynamics (MD) starting from a liquid phase with 
slow cooling rates have been used to generate a metastable phase (RS-type 
structure), however, no direct transition path was identified to link the 
proposed m-GST and a-GST phases. Thus, a new approach that connects the two 
phases at the atomistic level becomes highly desirable.  

\begin{figure}[t!]
\begin{center}
\scalebox{0.71}{\includegraphics{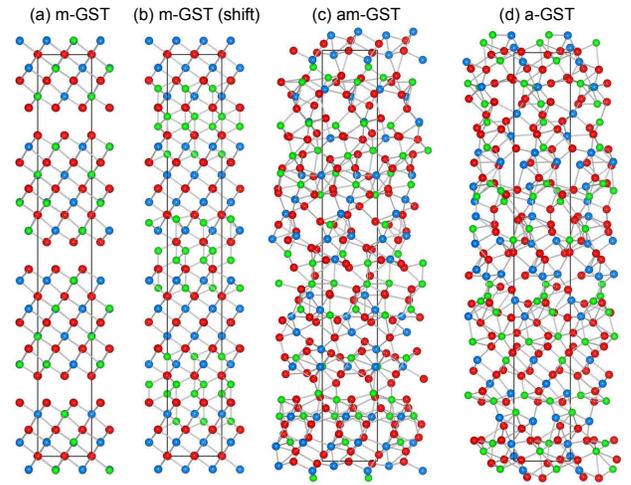}}
\caption{\label{gst_struct}
(Color online)
Structure models of the GST phases. (a) Metastable crystalline GST (m-GST). 
(b) m-GST (shift) structure, in which the Ge atoms occupy the 4-fold 
tetrahedron sites with lowest energy. (c) Amorphous GST obtained at zero 
temperature (am-GST) using modified m-GST structures (m-GST with Ge shift), in 
which the tetrahedron Ge sites were initially occupied. (d) Amorphous GST 
obtained by high temperature molecular dynamics DFT calculations (a-GST). The 
Ge, Sb, and Te atoms are indicated in green, blue, and red, respectively.
}
\end{center}
\end{figure}

In this work, using first-principles methods, we will address the following 
open questions: Is there a dominant structure link between both phases? What 
are the roles of GeTe and Sb$_2$Te$_3$ in GST? We will show in this Letter 
that combined static (zero temperature) and MD (high-temperature) 
first-principles calculations can explain the phase transition mechanism 
between the m-GST and a-GST phases. Moreover, our study shows that generating 
the amorphous phase from a known crystalline phase provides a better 
understanding of the structural relationship between both phases. Thus, it 
provides a new avenue for further study of amorphous materials phase change 
transitions.  

Our static total energy and MD calculations are based on the all-electron 
projected augmented wave (PAW) method\cite{Blochl-1994-17953,Kresse-1999-1758} 
and density functional theory (DFT) within the generalized gradient 
approximation (GGA-PBE)\cite{Perdew-1996-3865} as implemented in VASP.
\cite{Kresse-1993-13115,Kresse-1996-11169} To represent the metastable phase 
(RS-type structure), we employ a hexagonal $(2{\times}2{\times}1)$ unit cell, 
in which the Te atoms are stacked along of the [0001] direction.
\cite{DaSilva-2008-224111} The MD calculations were performed employing cubic 
and hexagonal cells with 108 to 126 atoms. The total energies and equilibrium 
volumes for all structures in both crystalline and amorphous phases were 
obtained by full relaxation of the volume, shape, and atomic positions of the 
unit cell to minimize the quantum mechanical 
stresses and forces. 

To understand the phase transition, we first established the crystal structure 
of the m-GST phase,\cite{DaSilva-2008-224111} as shown in 
Fig.~\ref{gst_struct}. The obtained structure is consistent with experimental 
results and provides new insights into m-GST.\cite{Yamada-2000-7020,Matsunaga-2004-104111,Park-2005-093506,Kolobov-2004-703,Matsunaga-2007-161919}
In this layered-structure the ordered intrinsic vacancies separate the 
building block units (GST), in which the Ge and Sb atoms are intermixed in 
planes. All Ge atoms are 6-fold coordinated in m-GST. However, it has been 
reported that up to one fifth of the Ge atoms are 4-fold coordinated with Te 
atoms in a-GST (GeTe$_4$), while the remaining Ge atoms form 4-fold motifs 
with combined Ge, Sb, and Te atoms.\cite{Jovari-2008-035202} We notice that 
tetrahedral Ge atoms can be obtained by shifting the octahedral Ge atoms in 
m-GST along the hexagonal $c$ direction, i.e., there are two tetrahedron sites 
for each Ge atom, Fig.~\ref{gst_struct}. In order to identify the lowest 
energy tetrahedron sites, we calculated the energetics for the occupation of 
each site by Ge atoms. The lowest energy sites are located in the intrinsic 
vacancy regions, while the highest energy sites are located in the center of 
the GST building blocks, i.e., there is a strong preference for the four-fold 
Ge atoms to be located in or near intrinsic vacancy regions. Assuming that all 
the Ge atoms are shifted from their octahedron sites and occupy the lowest 
energy tetrahedron sites, we find that 50\% of Ge will shift from the 
octahedra to tetrahedra along the [0001] direction, while the remaining 50\% 
shift along of the opposite direction. The m-GST (shift) structure in which 
all Ge atoms occupy the tetrahedral sites according to the distribution of 
intrinsic vacancies and energy barriers is shown in Fig.~\ref{gst_struct}b, 
which leads to the formation of Ge$-$Ge bonds. This configuration is highly 
unstable, and the system will relax without energy barrier to a lower energy 
phase (see am-GST structure in Fig.~\ref{gst_struct}c). 

\begin{figure}[t!]
\begin{center}
\scalebox{0.30}{\includegraphics{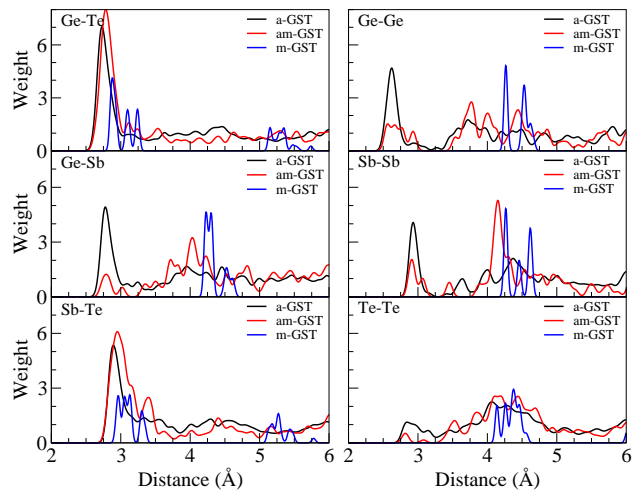}}
\caption{\label{MD_vs_T=0}
Pair-correlation functions of various GST phases. Amorphous GST obtained by 
molecular dynamic calculations (a-GST, black lines). Amorphous GST obtained 
from occupation of tetrahedron sites in m-GST and complete relaxation (am-GST, 
red lines). Meta-stable GST phase (m-GST, blue lines, scaled by 0.50).} 
\end{center}
\end{figure}

\begin{table}[b!]
\caption{\label{lattice_parameters}
Bond lengths (in {\AA}) of Ge$_2$Sb$_2$Te$_5$ (GST) in the amorphous and 
crystalline phases.  
}
\begin{ruledtabular}
\begin{tabular} {lccccc}
        & \multicolumn{3}{c}{Amorphous GST} & \multicolumn{2}{c}{Meta-stable GST} \\ \cline{2-4}\cline{5-6}
        & a-GST & am-GST & Exp. & m-GST & Exp. \\ \hline 
Ge$-$Te & 2.74 & 2.79 & $2.60-2.63$\footnotemark[1] & $2.87-3.24$ & $2.83-3.15$\footnotemark[2] \\
Sb$-$Te & 2.91 & 2.96 & $2.82-2.85$\footnotemark[1] & $2.96-3.30$ & 2.91\footnotemark[2] \\
Te$-$Te & 4.16 & 4.28 &      & $4.14-4.38$ & 4.26\footnotemark[2] \\ 
Ge$-$Ge & 2.63 & 2.64 & $2.47-2.48$\footnotemark[1] & $4.27-4.62$ &      \\ 
Ge$-$Sb & 2.79 & 2.78 & 2.69\footnotemark[1] & $4.23-4.53$ &      \\ 
Sb$-$Sb & 2.93 & 2.92 &      & $4.27-4.62$ &     \\
\end{tabular}
\end{ruledtabular}
{
\footnotetext[1]{Exp. Reference~\onlinecite{Jovari-2008-035202,Kolobov-2004-703,Baker-2006-255501}.}
\footnotetext[2]{Exp. Reference~\onlinecite{Kolobov-2004-703}.}
}
\end{table}

To provide a more direct structural link between the m-GST and a-GST phases, 
we first generated a-GST structures using first-principles MD simulations at 
high temperatures, $T$, using the same approach adopted in previous a-GST 
studies.\cite{Akola-2007-235201,Sun-2007-055505,Caravati-2007-171906,Hegedus-2008-399}
Secondly, we generated several amorphous structures from modified m-GST 
structures, in which a percentage of the Ge atoms (100\%, 75\%, 50\%, 25\%) 
are shifted to the tetrahedron sites from the lower energy octahedron sites 
(m-GST with shift Ge). The goal is to show that amorphous structures obtained 
in this way (am-GST in Fig.~\ref{gst_struct}) can preserve most of the 
structural features present in the a-GST generated by conventional MD 
calculations (a-GST in Fig.~\ref{gst_struct}), and therefore provide a direct 
structural link and solid evidence to support the mechanisms that determines 
the phase transition from m-GST to a-GST. The amorphous structures obtained by 
both approaches are shown in Fig.~\ref{gst_struct}. To quantify our analysis, 
we calculated the pair correlation (PC) functions, which are shown in 
Fig.~\ref{MD_vs_T=0}. For the a-GST structures, the PC functions were averaged 
over five structures, while the PC functions of the am-GST structures were 
calculated for ten structures with different initial occupation of the Ge 
tetrahedron sites; the structure that provided the best agreement with the 
a-GST PC functions is shown in Fig.~\ref{MD_vs_T=0}. 

\begin{figure}[t!]
\begin{center}
\scalebox{0.27}{\includegraphics{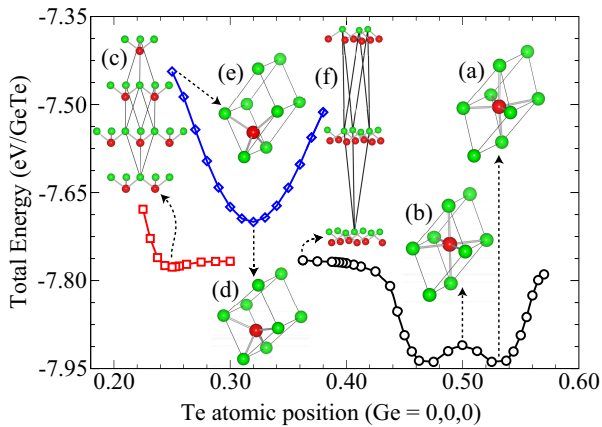}}
\caption{\label{GeTe_path}
Potential energy path for atomic displacements of Ge atoms along of the 
rocksalt (RS) [111] direction of GeTe. (a) Distorted RS structure. (b) Perfect 
RS structure. (c) Long Ge$-$Te bonds zincblende (ZB) structure. (d) Graphite 
like structure. (e) Perfect ZB structure. (f) RS-layer structure. 
}
\end{center}
\end{figure}

Our PC function analysis shows that for all the am-GST structures, the one in 
which 50\% of the Ge atoms are shifted from the octahedron to the tetrahedron 
sites along the hexagonal [0001] direction and the rest 50\% moves along the 
$[000\bar{1}]$ direction reproduces almost all features present in the PC 
functions of a-GST, although some minor differences still exist. Furthermore, 
even minor features are well-described by both structures, with the formation 
of Ge$-$Ge bonds and cavity regions, both of which have been identified as key 
characteristics of a-GST.\cite{Baker-2006-255501,Akola-2007-235201,Caravati-2007-171906,Hegedus-2008-399} 
We observe that am-GST structures in which less than half of the Ge atoms are 
moved to the tetrahedron sites do not yield PC functions similar to the a-GST
structures, instead they show strong similarity to the PC function calculated
for m-GST (see Fig.~\ref{MD_vs_T=0}). Furthermore, we observed that only the 
am-GST structures in which the Ge atoms initially occupy four-fold sites in or 
near the intrinsic vacancies lead to structure properties in good agreement 
with the calculated MD a-GST structures. Thus, it suggests that the location 
of the intrinsic vacancies plays an important role in the phase transition, 
which can be explained by the lower energy barriers for Ge displacements close 
to intrinsic vacancy regions. For the lowest energy m-GST structures, in which 
the intrinsic vacancies are ordered in a plane perpendicular to $c$. However, 
at high temperature or under non-equilibrium growth conditions the intrinsic 
vacancies may distribute more randomly among the cation sites, which is 
expected to play an important role in the pattern of shifted Ge atoms from 
their stable octahedra.

Our predicted results are in good agreement with available experimental data. 
For example, using the calculated equilibrium volumes for both phases, we 
obtained a density of 5.89~g/cm$^3$ (m-GST) and 5.35~g/cm$^3$ (a-GST and 
am-GST), i.e. the amorphization gives rise to a volume expansion, which 
decreases the density by about 9.20\%. The experimentally observed expansion 
is on the order of 6.4\%.\cite{Njoroge-2002-230} The volume expansion upon 
amorphization is a consequence of the Ge atoms moving to the lower 
coordination sites in the a-GST structures. Therefore, the smaller volume 
deformation observed in the experimental sample may indicate that the 
amorphization process in not complete,
\cite{Yamada-1991-2849,Jovari-2008-035202}  i.e. not all the Ge atoms are 
moved away from their stable octahedron sites.


Comparison of the total energies reveals that the a-GST structure is about 
$140-182$~meV/atom higher in energy than the lowest energy m-GST structure, 
which corresponds to the energy limit between the fully amorphized (100\% 
shift of the Ge atoms) and the ordered m-GST structure. Differential scanning 
calorimetry measurements obtained $28-42$~meV/atom.\cite{Kalb-2003-2389} We 
found that the calculated energy differences decrease by about 30~meV/atom if 
the intrinsic vacancies become disordered in m-GST. Furthermore, the energy 
difference could be much smaller (e.g. about 50~meV/atom) if only a fraction 
of the Ge atoms undergo site transitions. This again suggests that full scale 
amorphorization of GST or a complete ordering of Ge, Sb, and intrinsic 
vacancies in m-GST may not be typical in the GST phases.  


The averaged bond lengths calculated for both phases are summarized in 
Table~\ref{lattice_parameters} along with available experimental results.
\cite{Jovari-2008-035202,Kolobov-2004-703,Baker-2006-255501} 
The calculated bond lengths deviate by about $3 - 6$\% from the experimental 
results; however, most of the error is due to the use of GGA in our 
calculations, which systematically overestimates the lattice constants by 
about 3\%. Furthermore, it is important to notice that the nearest-neighbor 
distances are spread over a large range of values, e.g.  Ge$-$Te is from 2.67 
to 2.94~{\AA} and Sb$-$Te is from 2.86 to 3.23~{\AA}. We found a contraction 
in the averaged Ge$-$Te bond lengths in a-GST of up to 10\% compared with 
m-GST, e.g. Ge$-$Te decreases from $2.87-3.24$~{\AA} (m-GST) to 
$2.67-2.94$~{\AA} (a-GST), while experimental measurements obtained a decrease 
of about 12\%. Similar trends exist for Sb$-$Te. 

To understand the relaxation effects introduced by the shift of Ge atoms from 
octahedron to tetrahedron sites, we calculated the potential energy path along 
the RS [111] direction for GeTe as a function of Ge shift from octahedron 
(perfect RS) to the tetrahedron (zincblende) sites. The results are shown in 
Fig.~\ref{GeTe_path}. As expected, the distorted RS structure has the lowest 
energy (54~meV lower than the perfect RS structure), in which the distortion 
is driven by Peierls-type level repulsion near the band edge. Unexpectedly, 
the zincblende (ZB) structure in which the Ge atoms occupy the tetrahedron 
sites with bond angles (Ge$-$Te$-$Ge) of $109.47^{\circ}$, is not a local 
minimum as would be expected based on the general trends for binary 
semiconductors. In fact, we found that the ZB structure relaxes without energy 
barrier to the `graphite-like' or to the `long-Ge$-$Te' structures, which have 
lower energies than the high-symmetry ZB phase. Thus, Ge at ideal tetrahedral 
sites are intrinsically unstable in GeTe, which drives the Ge atoms at 
tetrahedral sites in GST to move away and adopt a variety of lower symmetry 
coordination environments. 

The variety of coordination environments found in the GeTe energy surface is 
remarkable. From Ge site occupation of (0.25,0.25,0.25) to (0.40,0.40,0.40), 
three structures have similar energies, i.e. `long-Ge$-$Te', layered-ZB, and 
layered-RS. In `long-Ge$-$Te', the Ge atoms form three short bonds 
(2.77~{\AA}) and one {\it long} bond (4.68~{\AA}) with the Te atoms. However, 
Ge is only three-fold coordinated in the layered structures with bond lengths 
of 2.76~{\AA}, which is 2.90\% (14.51\%) smaller than the short (longer) 
Ge$-$Te bond lengths in the distorted RS structure. As the layered GeTe 
structures are lower in energy than the graphite-like phase and only about 
100~meV/f.u. higher than the distorted RS structure, it indicates a strong 
tendency of Ge atoms to form four-fold motifs with three short Ge$-$Te bonds 
(about 2.76~{\AA}) and bond angles of about $90^{\circ}$. Similar results, 
e.g. short bond lengths and average bond angles of about $90^{\circ}$, are 
observed by our calculations for a-GST, which is also consistent with previous 
MD results for a-GST,\cite{Akola-2007-235201,Caravati-2007-171906,Hegedus-2008-399} 
as well as by experimental observations.\cite{Kolobov-2004-703,Kolobov-2006-035701,Jovari-2008-035202}  
Therefore, the inherent instability of Ge at the tetrahedral sites, low 
displacement energy, and unique coordination preferences of GeTe plays an 
important role in the formation of a-GST.

In summary, using first-principles calculations, we obtained a direct 
structural link between the meta-stable and amorphous GST phases, as well as 
the role of the parent compounds. The Sb$_2$Te$_3$ provides intrinsic lattice 
vacancies, while GeTe contributes its RS-type structure in which Ge 
displacements along the RS [111] direction can be realized at low energy cost. 
The instability at the tetrahedral sites leads to the generation of disordered 
GST structures in which the Ge atoms are mostly four-fold coordinated with 
three short Ge$-$Te bond lengths. As the displacement has the lowest energy 
near intrinsic vacancy sites, our analysis suggests that a high degree of 
amorphization can be achieved most easily when the system has a composition of
(GeTe)$_2$(Sb$_2$Te$_3$), i.e., is consistent with the observation that GST 
has the highest figure of merit of all Ge$-$Sb$-$Te compounds. Moreover, we 
show that generating amorphous materials directly from its crystalline 
counterpart provides a better approach to understand these type of phase 
transitions present in phase change materials. 


\end{document}